\documentclass[conference,10pt]{IEEEtran}
\usepackage{graphicx}
\usepackage{verbatim}
\usepackage{upgreek}
\usepackage{amssymb,amsmath}
\usepackage{color}
\usepackage{epstopdf}
\usepackage{bm}
\usepackage{cite}
\usepackage{array,color}
\usepackage{algorithm}
\usepackage{algorithmicx}
\usepackage{algpseudocode}
\usepackage{amsmath}
\usepackage{amsfonts}
\usepackage{amsthm}
\usepackage{graphics}
\usepackage{epsfig}
\usepackage{soul}
\usepackage{booktabs}
\usepackage{multirow}
\soulregister\cite7
\soulregister\ref7
\usepackage{subfigure}  
\usepackage{tabularx}
\usepackage{makecell}
\usepackage{graphicx}
\usepackage[font=small]{caption}

\usepackage{etoolbox}
 
\makeatletter
\patchcmd{\@makecaption}
{\scshape}
{}
{}
{}
\makeatletter
\patchcmd{\@makecaption}
{\\}
{.\ }
{}
{}
\makeatother

\ifCLASSINFOpdf
\else
\fi

\begin{document}

\title{Computation Capacity Maximization for Pinching Antennas-Assisted  Wireless Powered MEC Systems}

\author{
	\IEEEauthorblockN{Peng Liu\IEEEauthorrefmark{1},~Meng Hua\IEEEauthorrefmark{2},~Guangji Chen\IEEEauthorrefmark{3},~Xinyi Wang\IEEEauthorrefmark{1},~Zesong Fei\IEEEauthorrefmark{1}}
	
	\IEEEauthorblockA{\IEEEauthorrefmark{1}School of Information and Electronics, Beijing Institute of Technology (BIT), Beijing, China\\
	\IEEEauthorrefmark{2}Department of Electrical and Electronic Engineering, Imperial College London, London, U.K\\
\IEEEauthorrefmark{3}School of Electronic and Optical Engineering, Nanjing University of Science and Technology, Nanjing, China}

	Email: bit\_peng\_liu@163.com,~m.hua@imperial.ac.uk,~guangjichen@njust.edu.cn,~\{wangxinyi, feizesong\}@bit.edu.cn
}

\maketitle

\begin{abstract}

In this paper, we investigate a novel wireless powered mobile edge computing (MEC) system assisted by pinching antennas (PAs), where devices first harvest energy from a base station and then offload computation-intensive tasks to an MEC server. As an emerging technology, PAs utilize long dielectric waveguides embedded with multiple localized dielectric particles, which  can be spatially configured through a pinching mechanism to effectively reduce large-scale propagation loss. This capability facilitates both efficient downlink energy transfer and uplink task offloading. To fully exploit these advantages, we adopt a non-orthogonal multiple access (NOMA) framework and formulate a joint optimization problem to maximize  the system’s computational capacity by  jointly optimizing device transmit power, time allocation, PA positions in both uplink and downlink, and radiation control. To address the resulting non-convexity caused by variable coupling, we develop an alternating optimization algorithm that integrates particle swarm optimization (PSO) with successive convex approximation. Simulation results demonstrate that the proposed PA-assisted design substantially improves both energy harvesting efficiency and computational performance compared to conventional antenna systems.

\vspace{1ex}
\end{abstract}

\IEEEpeerreviewmaketitle
\section{Introduction}
With the rapid proliferation of the Internet of Things (IoT), an enormous number of smart devices are being deployed across diverse scenarios, such as environmental monitoring, smart homes, and intelligent healthcare \cite{iot}. However, constrained by manufacturing costs and form-factor limitations, most IoT devices are equipped with low-capacity batteries and resource-limited processors, making it difficult to support latency-sensitive and computation-intensive applications. To overcome these challenges, wireless powered mobile edge computing (MEC) has emerged as a promising solution by combining wireless power transfer (WPT) with MEC \cite{MEC-WPT}. While WPT provides sustainable energy replenishment for battery-limited devices, MEC offers nearby high-performance computing resources to offload intensive tasks. This integration effectively addresses the dual bottlenecks of energy supply and computational capability, greatly enhancing the scalability and intelligence of IoT systems. However, despite its potential, the performance of wireless powered MEC remains fundamentally limited by wireless propagation loss, which degrades both energy transfer efficiency and task offloading capability.

Recent advances in flexible antenna systems, such as reconfigurable intelligent surfaces (RIS) \cite{RIS1, RIS2, RIS3}, fluid antennas \cite{fas1, fas2}, and movable antennas \cite{Mov1, Mov2, Mov3}, have opened up new opportunities for enhancing wireless channel conditions. RISs adjust the amplitude and phase of reflected signals via tunable electromagnetic elements, enabling intelligent propagation control. However, their performance is often limited by the double-fading effect caused by cascaded base station (BS)-RIS-device links. Fluid and movable antennas improve link quality by physically relocating the antenna, but their movement is typically constrained to within a few wavelengths, limiting their effectiveness in mitigating large-scale path loss.

To address these limitations, pinching antennas (PAs) were proposed by NTT DOCOMO in 2022 as a novel solution for highly reconfigurable antenna deployment \cite{suzuki2022pinching}. Drawing inspiration from leaky-wave antenna principles, PAs employ long dielectric-filled waveguides, typically tens of meters in length, embedded with localized ``pinched” dielectric particles that enable controllable signal transmission or reception \cite{PASS0}. \textcolor{black}{By selectively positioning dielectric elements along the waveguide}, PAs can flexibly steer radiation to establish favorable line-of-sight (LoS) channels to nearby devices. Unlike traditional flexible antenna systems, PA waveguides can be arbitrarily extended, allowing effective apertures to be positioned closer to devices, thereby significantly reducing path loss \cite{PASS4}. In addition, PAs require only  the simple addition or removal of dielectric materials at designated points, offering a scalable and low-cost implementation. The fundamental principles and future potential of PA systems have been introduced in \cite{PASS1}, highlighting their promise for next-generation reconfigurable wireless infrastructure. Building on this foundation, \cite{PASS2} and \cite{PASS3} investigated PA-assisted downlink and uplink communication systems, respectively, and demonstrated significant data rate improvements over traditional antenna arrays. However, their application in wireless powered MEC systems remains largely unexplored. 

Motivated by these findings, we investigate a novel wireless powered MEC system enhanced by PAs, where devices harvest energy from downlink WPT and offload computation tasks in the uplink. By \textcolor{black}{configuring} localized dielectric particles along long waveguides, PAs enable flexible signal transmission and reception while  effectively reducing path loss, thereby improving both energy harvesting and task offloading efficiency. To fully leverage this capability, we adopt a non-orthogonal multiple access (NOMA) scheme for simultaneous offloading and formulate a joint optimization problem involving device transmit power, time allocation, PA positions in both phases, and radiation control. To tackle this non-convex problem, we develop an alternating optimization framework that integrates particle swarm optimization (PSO) with successive convex approximation (SCA). Simulation results verify that the proposed PA-assisted design significantly outperforms conventional antenna systems in both energy and computation performance.

\setlength{\abovecaptionskip}{1pt} 
\section{System Model  and Problem Formulation}

\begin{figure}[!t]
	\centering
	\includegraphics[width=3.1in]{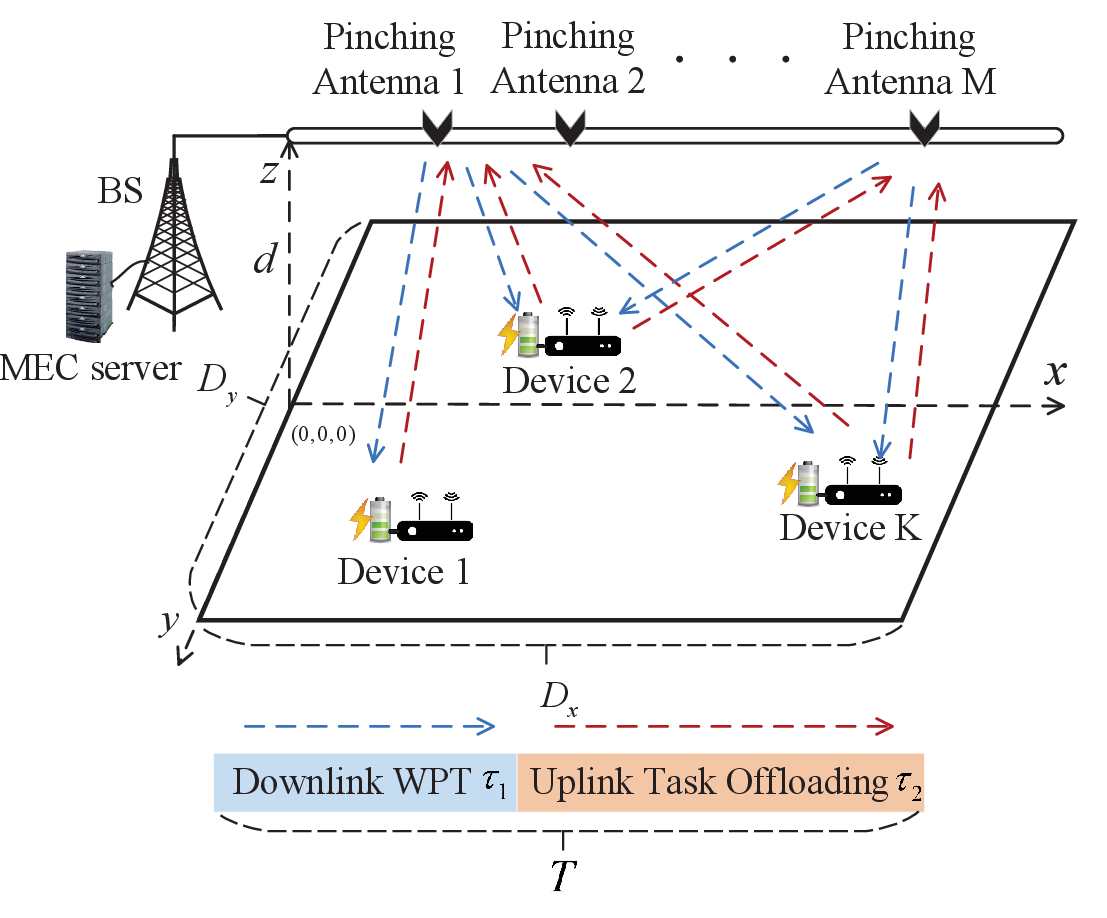}
	\caption{Illustration of PA-assisted  wireless powered MEC systems.}
	\label{fig:sys}
\end{figure}
As shown in Fig. 1, we consider a PA-assisted wireless powered MEC system consisting of a BS equipped with $M$ PAs and an MEC server, serving  $K$ single-antenna devices. The devices are assumed to be randomly distributed within a rectangular area of size $D_x \times D_y$, and their positions are denoted by $\bm{\Psi}_k=(x_k, y_k, 0), k=1,\cdots,K$. The waveguide connected to the BS is deployed along the $x$-axis at a height of $d$, with its feed point located at a  reference position $\bm{\Psi}^P_0=(0, 0, 0)$. The PAs are mounted along  the waveguide and can be spatially configured at arbitrary positions. The location of the $m$-th PA is denoted by $\bm{\Psi}^P_m=(x_m, 0, d)$. 

We assume that each device is equipped with a rechargeable battery and an energy harvesting circuit, enabling it to execute local computation or offload tasks by harvesting energy from the BS's broadcast signals. Moreover, a time-division duplexing (TDD) protocol is employed to separate downlink WPT phase and uplink task offloading phase. Specifically, each transmission frame $T$ is divided into two phases as shown in Fig. 1: $\tau_1$ for WPT and $\tau_2$ for offloading. During the WPT phase, the downlink channel from the feed point, through the $m$-th PA, to the $k$-th device is given by \cite{PASS2}
\begin{equation}
h^D_{mk}=\underbrace{\frac{\eta{\alpha }_{m}}{\left\|{\bm{\Psi}_k}-\bm{\Psi}^{PD}_m\right\|}}_{\text {free-space path loss }} \cdot \underbrace{e^{-j \frac{2 \pi}{\lambda}\left\|{\bm{\Psi}_k}-\bm{\Psi}^{PD}_m\right\|}}_{\text {free-space phase shift }} \cdot \underbrace{e^{-j \frac{2 \pi}{\lambda_g}\left\|\bm{\Psi}^P_0-\bm{\Psi}^{PD}_m\right\|}}_{\text {in-waveguide phase shift }},
\end{equation}
where $\eta=\frac{c}{4\pi f_c}$, with $c$ and $f_c$ denoting the speed of light and carrier frequency, respectively.  $\lambda$  is the free-space wavelength, and  $\lambda_g=\frac{\lambda}{n_e}$ is the guided wavelength with $n_e$ being the effective refractive index of the dielectric waveguide \cite{PASS2}. \textcolor{black}{$\alpha_m$ represents the radiation factor of the $m$-th pinching antenna, which is governed by the effective coupling length introduced when an external dielectric is clamped onto the waveguide. This factor captures how efficiently energy leaks from the waveguide into free space at the pinching location, and is subject to the normalization constraint $\sum_{m=1}^{M} \alpha_m^2 \leq 1$
to ensure total radiated power remains bounded.} $\bm{\Psi}^{PD}_m$ is the position of $m$-th PA during the downlink WPT phase. Moreover, compared to free-space propagation, waveguide propagation exhibits extremely low attenuation (e.g., 0.01–0.03 dB/m at 15 GHz for a circular copper waveguide), and thus, the propagation  loss along the waveguide is assumed to be negligible in this work \cite{PASS1}. Based on the linear energy harvesting model in \cite{WPT2}, the energy harvested by device $k$ can be expressed as\footnote{\textcolor{black}{Although non-linear energy harvesting is more accurate, it is approximately equivalent to the linear model under low-energy conditions. We adopt the linear model to simplify the analysis of PA impact, and note that our framework remains applicable to non-linear models as well.}}
\begin{equation}
{{E}_{k}}=\beta {{\tau }_{1}}{{P}_{B}}{{\left| \sum_{m=1}^{M}{h^D_{mk}} \right|}^{2}},\forall k,
\end{equation}
where $\beta\in(0,1]$ denotes the energy conversion efficiency and $P_B$ is the transmit power of BS.

For the uplink offloading phase, we assume that the devices adopt a partial offloading strategy, in which each computation task is divided into two parts: one is executed locally, and the other is offloaded to the MEC server.   Let $f_k$ denote the computational frequency (in cycles/s) of device $k$, the number of bits computed locally and the corresponding computation energy consumption are respectively given by
\begin{equation}
C^L_k = \frac{Tf_k}{D},\quad~e_k = T\kappa f^3_k, ~~~\forall k,
\end{equation}
where $D$ (in cycles/bit) denotes the task computation intensity, and $\kappa$ is the  power coefficient related to the chip architecture \cite{WPT2}. The uplink channel from the $k$-th device,  via the $m$-th PA, to the feed point,  is modeled as
\begin{equation}
h^U_{mk}=\frac{\eta e^{-j \frac{2 \pi}{\lambda}\left\|{\bm{\Psi}_k}-\bm{\Psi}^{PU}_m\right\|} \cdot e^{-j \frac{2 \pi}{\lambda_g}\left\|\bm{\Psi}^P_0-\bm{\Psi}^{PU}_m\right\|}}{\left\|{\bm{\Psi}_k}-\bm{\Psi}^{PU}_m\right\|},
\end{equation}
where $\bm{\Psi}^{PU}_m$ is the position of $m$-th PA during the uplink offloading phase.  Assuming that all devices transmit simultaneously over the same frequency band using NOMA, the total number of offloaded bits within a transmission frame is expressed as
\begin{equation}
{C^{\text{off}}}=B{{\tau }_{2}}{{\log}_{2}}\left( 1+\frac{\sum_{k=1}^{K}{{p}_{k}{{\left| \sum_{m=1}^{M}h^U_{mk} \right|}^{2}}}}{M{\sigma }_b^{2}} \right),\quad \forall k,
\end{equation}
where $B$ denotes the signal bandwidth, $p_k$ is the transmit power of device $k$, ${\sigma }_b^{2}$ represents the power of the additive white Gaussian noise. Consequently, the computation capacity of the system can be expressed as
\begin{equation}
 C^{\text{sum}}={C^{\text{off}}}+\sum_{k=1}^{K}C^L_k.
\end{equation}

In addition,  we can observe that  the system’s computation capacity is influenced by the positions of the PAs in both downlink and uplink phases $\bm{\Psi}^{PD}_m$ and $\bm{\Psi}^{PU}_m$, as well as the radiation factors $\alpha_m$. By flexibly \textcolor{black}{configuring} the PAs, the path loss between BS and the devices can be effectively reduced, thereby improving energy harvesting efficiency and facilitating higher task offloading rates. Furthermore, optimizing the device transmit power $p_k$, time allocation $\{\tau_1, \tau_2\}$, and local computation frequency $f_k$ can further enhance energy efficiency and thereby improve the overall computation capacity. Therefore, we aim to jointly optimize $\bm{\Psi}^{PD}_m$, $\bm{\Psi}^{PU}_m$, $\alpha_m$, $p_k$, $\tau_1$, $\tau_2$, and  $f_k$ to maximize the system computation capacity. The corresponding optimization problem is formulated as 
\begin{align} 
\label{prob29}
\max_{\substack{\bm{\Psi}^{PD}_m, \bm{\Psi}^{PU}_m, \alpha_m,\\p_k, \tau_1, \tau_2, f_k}} & ~C^{\text{sum}}\\
\text{s.t.}~~\quad
&|\bm{\Psi}^{PD}_m -\bm{\Psi}^{PD}_{m'}| \geq \Delta,~ \forall m, m', m\neq m', \tag{\ref{prob29}a}\\
& |\bm{\Psi}^{PU}_m -\bm{\Psi}^{PU}_{m'}| \geq \Delta,~ \forall m, m', m\neq m', \tag{\ref{prob29}b}\\
& x_m\in[0, D_x],~~\forall m,\tag{\ref{prob29}c}\\
& p_k\tau_2+e_k\leq E_k, ~~\forall k, \tag{\ref{prob29}d}\\
& \sum_{m=1}^{M}\alpha^2_m\leq 1,  \tag{\ref{prob29}e} \\
& \tau_1+\tau_2 \leq T,  \tag{\ref{prob29}f}
\end{align}
where constraints (\ref{prob29}a) and (\ref{prob29}b) impose  a minimum spacing $\Delta$ between adjacent PAs to mitigate coupling effects, constraint (\ref{prob29}c) ensures that all PA positions lie  within the range of the waveguide. \textcolor{black}{In addition, constraint (7d) guarantees that each device’s total energy consumption, including transmission and local computation, does not exceed the harvested energy.}

\section{Proposed Algorithm}
Due to the presence of PA position variables in both the numerator and denominator of the objective function and the energy expressions $E_k$, as well as the intrinsic coupling among $\bm{\Psi}^{PU}_m$, $p_k$, and $\tau_1$, problem (\ref{prob29}) is  non-convex and difficult to solve. In this section, we develop an efficient alternating optimization algorithm to tackle this problem.
\subsection{Uplink PA position Optimization}
We first optimize the uplink PA position $\bm{\Psi}^{PU}_m$  given $\bm{\Psi}^{PD}_m$, $\alpha_m$, $p_k$, $\tau_1$, $\tau_2$ and  $f_k$. The corresponding optimization subproblem  can be reformulated as 
\begin{align} 
\label{prob23}
\max_{\bm{\Psi}^{PU}_m} &  \quad \sum_{k=1}^{K}{{p}_{k}{{\left| \sum_{m=1}^{M}h^U_{mk} \right|}^{2}}} \\
\text{s.t.}~&~~~ \text{(\ref{prob29}\text{b})},\text{(\ref{prob29}\text{c})}. \notag
\end{align}

Due to the non-convexity of the objective function and non-convex constraint (7b), problem (\ref{prob23}) is still challenging to solve using conventional convex optimization tools. To address this, we employ the PSO algorithm,  which is an effective global search method with low computational complexity suitable  for such problems \cite{PSO}. To facilitate its application, problem (\ref{prob23}) is transformed into the following unconstrained form:
\begin{align} 
\label{prob24}
\min_{\bm{\Psi}^{PU}_m} &   -\sum_{k=1}^{K}{{p}_{k}{{\left| \sum_{m=1}^{M}h^U_{mk} \right|}^{2}}}+\mathbb{I}(\bm{\Psi}^{PU}_m),
\end{align}
where $\mathbb{I}(\bm{\Psi}^{PU}_m)$ is the penalty term associated with constraints (\ref{prob29}b) and (\ref{prob29}c), defined as follows
\begin{equation}
\mathbb{I}(\bm{\Psi}^{PU}_m)= \begin{cases}0, & \text { if (\ref{prob29}b) and (\ref{prob29}c) hold,}  \\ +\infty, & \text { otherwise. }\end{cases}
\end{equation}

Next, we employ the PSO algorithm to solve problem (\ref{prob24}). We first randomly initialize $N$ particles $\mathbf{x}_n=\{x_{1,n},\cdots,x_{M,n}\}, n=1,...,N$ within the feasible search space. During the optimization process, each particle evaluates its fitness value based on the objective function of (\ref{prob24}), denoted by $f(\mathbf{x}_n)$, and subsequently updates its personal best position $\bm{pbest}_n$. The global best position $\bm{gbest}$ is then determined by comparing the fitness values across all particles. \textcolor{black}{These positions collectively guide the swarm's exploration of the solution space in search of an optimal or near-optimal solution.} At the $r$-th iteration, the position and velocity of each particle are updated as 
\begin{equation}\label{prob1}
\begin{aligned}
& \mathbf{x}_{n}^{r}=\mathbf{x}_{n}^{r-1}+\mathbf{v}_n^{r},\\& \mathbf{v}_n^{r}=b_0 \mathbf{v}_n^{r-1}+b_1 c_1\left(\bm{pbest}_n-\mathbf{x}_{m,n}^{r-1}\right)+b_2 c_2\left(\bm{gbest}-\mathbf{x}_{m,n}^{r-1}\right), 
\end{aligned}
\end{equation}
where $b_0$ is the inertia weight, which controls the momentum of the particle, $b_1$ and $b_2$ are learning coefficients, $c_1$ and $c_2$ are random variables in the range [0,1]. ${v}_n^{r}$ and ${x}_{m,n}^{r}$ denote the velocity and position of the $n$-th particle in the $r$-th iteration,  respectively.  The iterations are repeated until the fitness value converges, yielding the optimized uplink PA positions $\bm{\Psi}^{PU}_m$. For clarity, the detailed PSO procedure is presented in Algorithm 1.

\subsection{Downlink PA position Optimization}
For the optimization of downlink PA positions $\bm{\Psi}^{PD}_m$, we can be observed that constraint (\ref{prob29}d) is satisfied with equality at the optimal solution. This is because  each device tends to fully utilize its harvested energy to maximize its computation capacity. Therefore, we have 
\begin{equation} \label{p}
p_k = \frac{1}{\tau_2}(E_k - e_k).
\end{equation}

 Given fixed  $\bm{\Psi}^{PU}_m$, $\alpha_m$, $p_k$, $\tau_1$, $\tau_2$, and  $f_k$, and by ignoring constant terms, the optimization subproblem with respect to  $\bm{\Psi}^{PD}_m$ is formulated as
\begin{align} 
\label{prob25}
\max_{\bm{\Psi}^{PD}_m} &  \quad \sum_{k=1}^{K}{{g}_{k}\beta {{\tau }_{1}}{{P}_{B}}{{\left| \sum_{m=1}^{M}h^D_{mk} \right|}^{2}}} \\
\text{s.t.}~&~~~ \text{(\ref{prob29}\text{a})},\text{(\ref{prob29}\text{c})}. \notag
\end{align}
where $g_k={{\left| \sum_{m=1}^{M}h^U_{mk} \right|}^{2}}$ denotes the uplink channel gain. We observe that problem (\ref{prob25}) exhibits a structure similar to  problem (\ref{prob24}) and thus  it can be efficiently solved using the PSO algorithm.

\renewcommand{\algorithmicrequire}{\textbf{Input:}}
\renewcommand{\algorithmicensure}{\textbf{Output:}}
\begin{algorithm}[t] 
	\caption{PSO Algorithm for Solving Problem (\ref{prob24})}
	\begin{algorithmic}[1] 
		\State \textbf{Initialize:} Randomly generate $N$  particles $\{\mathbf{x}_{n}\}, n=1,...,N$ within the feasible search space and initialize $r=1$, $b_0$, $b_1$, and $b_2$.
			\State Evaluate the fitness function $f(\mathbf{x}_n)$ for each particle.
			\State  Set $\bm{pbest}_n=\mathbf{x}^0_{n}$ and $\bm{gbest}$ =$\text{argmin}_{pbest_n}(f(\mathbf{x}_n))$
		\Repeat
		\For{$i = 1$ to $N$}
			\State Update position and velocity of $\mathbf{x}^{r-1}_{n}$ based on (\ref{prob1}).
		\If{$f(\mathbf{x}^{r}_n)\leq f(\bm{pbest}_n) $}
			\State $\bm{pbest}_n=\mathbf{x}^{r}_n$.
		\EndIf
		\If{$f(\mathbf{x}^{r}_n)\leq f(\bm{gbest}) $}
		\State $\bm{gbest}=\mathbf{x}^{r}_n$.
		\EndIf
		\State  $r=r+1$.
		\EndFor
		\Until {convergence.}
	\end{algorithmic}
\end{algorithm}
\subsection{Radiation Factors and  Device Transmit Power Optimization}
Next, we  optimize the radiation factors $\alpha_m$ and the transmission powers of the devices $p_k$. To facilitate the problem formulation, we define the radiation vector $\mathbf{w}=[\alpha_1,...,\alpha_m]^T$ \textcolor{black}{and construct the effective complex channel vector for each device $k$ as}
\begin{equation} 
\begin{aligned}
\small
\mathbf{u}_k = & \left[
\frac{e^{-j \frac{2 \pi}{\lambda}\left\|{\bm{\Psi}_k}-\bm{\Psi}^{PU}_1\right\| 
		- j \frac{2 \pi}{\lambda_g}\left\|\bm{\Psi}^P_0 - \bm{\Psi}^{PU}_1\right\|}}{\left\|{\bm{\Psi}_k}-\bm{\Psi}^{PU}_1\right\|}, \dots, \right. \\
&\qquad\qquad\qquad \left. \frac{e^{-j \frac{2 \pi}{\lambda}\left\|{\bm{\Psi}_k}-\bm{\Psi}^{PU}_M\right\| 
		- j \frac{2 \pi}{\lambda_g}\left\|\bm{\Psi}^P_0 - \bm{\Psi}^{PU}_M\right\|}}{\left\|{\bm{\Psi}_k}-\bm{\Psi}^{PU}_M\right\|}
\right]^T.
\end{aligned}
\end{equation}

Given (\ref{p}) \textcolor{black}{and fixing  $\bm{\Psi}^{PU}_m$, $\bm{\Psi}^{PD}_m$,  $\tau_1$, $\tau_2$, and  $f_k$, the optimization problem with respect to 
$\bf w$ can be reformulated as}
\begin{align} 
\label{prob26}
\max_{\mathbf{w}} &  \quad \sum_{k=1}^{K}{{g}_{k}\beta {{\tau }_{1}}{{P}_{B}}\eta^2{{\left|\mathbf{u}^T_k\mathbf{w} \right|}^{2}}} \\
\text{s.t.}~&~~~ \text{(\ref{prob29}\text{e})}. \notag
\end{align}

To further simplify the problem, we introduce auxiliary variables $\{t_k\},k=1,...,K$, and  reformulate  problem (\ref{prob26}) as 
\begin{align} 
\label{prob27}
\max_{\mathbf{w}} &  \quad \sum_{k=1}^{K}{{g}_{k}\beta {{\tau }_{1}}{{P}_{B}}\eta^2{t_k}} \\
\text{s.t.}~&~~~t_k\leq{\left|\mathbf{u}^T_k\mathbf{w} \right|}^{2}, \forall k,  \tag{\ref{prob27}a}\\&~~~ \text{(\ref{prob29}\text{e})}. \notag
\end{align}

However, (\ref{prob27}a) remains non-convex. To handle this  issue, we employ the SCA technique to linearize the constraint around a given point $\widehat{\mathbf{w}}$, yielding
\begin{equation} 
t_k \leq 2\widehat{\mathbf{w}}^{T}\Re(\mathbf{u}^*_k\mathbf{u}^T_k)\mathbf{w}-\widehat{\mathbf{w}}^{T}\Re(\mathbf{u}^*_k\mathbf{u}^T_k)\widehat{\mathbf{w}}, ~~\forall k,
\end{equation} 
where $\Re(\cdot)$ denotes the real part operator. Based on the above transformation, problem (\ref{prob27}) can be rewritten as
\begin{align} 
\label{prob30}
\max_{\mathbf{w}} &  \quad \sum_{k=1}^{K}{{g}_{k}\beta {{\tau }_{1}}{{P}_{B}}\eta^2{t_k}} \\
\text{s.t.}~&~~~t_k \leq 2\widehat{\mathbf{w}}^{T}\Re(\mathbf{u}^*_k\mathbf{u}^T_k)\mathbf{w}-\widehat{\mathbf{w}}^{T}\Re(\mathbf{u}^*_k\mathbf{u}^T_k)\widehat{\mathbf{w}}, ~~\forall k,  \tag{\ref{prob30}a}\\&~~~ \text{(\ref{prob29}\text{e})}. \notag
\end{align}

This reformulated problem is convex and can be efficiently solved using standard convex optimization solvers such as CVX \cite{2004Convex}. After obtaining the optimal radiation vector $\bf w$, the optimal  transmit power $p_k$ for each device can be obtained via (\ref{p}).

\subsection{Time Allocation and  Local Computational Frequency Optimization}
This subsection focuses on optimizing  the time allocation variables $\tau_1, \tau_2$,  and the local computational frequency $f_k$ with given PA positions  $\bm{\Psi}^{PU}_m$, $\bm{\Psi}^{PD}_m$, the radiation factors $\alpha_m$, and device transmit powers  $p_k$. The corresponding subproblem is formulated as
\begin{align} 
\label{prob28}
\max_{\tau_1, \tau_2, f_k} &  \quad {C^{\text{off}}}+\sum_{k=1}^{K}C^L_k \\
\text{s.t.}~&~~~ \text{(\ref{prob29}\text{d})},\text{(\ref{prob29}\text{f})}. \notag
\end{align}

\renewcommand{\algorithmicrequire}{\textbf{Input:}}
\renewcommand{\algorithmicensure}{\textbf{Output:}}
\begin{algorithm}[t] 
	\caption{Overall Algorithm for Solving Problem (\ref{prob29})}
	\begin{algorithmic}[1] 
		\State \textbf{Initialize:} Downlink PA positions $\bm{\Psi}^{PD}_m$, radiation factor vector $\widehat{\mathbf{w}}$, device transmit powers $p_k$, time allocation variables $\tau_1$, $\tau_2$, and local computation frequencies $f_k$.
		\Repeat
		\State  Optimize uplink PA positions $\bm{\Psi}^{PU}_m$ by solving problem (\ref{prob24}) using the PSO algorithm.
		\State  Optimize downlink PA positions $\bm{\Psi}^{PD}_m$ by solving problem (\ref{prob25}) using the PSO algorithm.
		\Repeat
		\State  Obtain the optimized radiation vector $\mathbf{w}^{*}$ by solving problem (\ref{prob27}).
		\State  Update $\widehat{\mathbf{w}}=\mathbf{w}^*$.
		\Until {convergence.}
		\State Compute the  optimal device transmit power $p^*_k$ via $p^*_k=\frac{1}{\tau_2}(E_k-e_k)$.
		\State  Optimize time allocation variables $\tau_1$, $\tau_2$, and local computation frequencies $f_k$ by solving problem (\ref{prob28}).
		\Until {convergence.}
	\end{algorithmic}
\end{algorithm}

It can be observed that problem (\ref{prob28}) is convex and can be solved using standard convex optimization tools such as CVX.

\subsection{Overall Algorithm and Convergence Analysis
}
We summarize the alternating optimization algorithm described in the previous subsections as Algorithm 2. Specifically, we first optimize the PA positions $\bm{\Psi}^{PU}_m$ and $\bm{\Psi}^{PD}_m$ for the uplink and downlink phases using the PSO algorithm. Next, the radiation factor vector $\mathbf{w}$ is updated iteratively via the SCA technique, based on which the optimal transmit power $p_k$ is determined. Finally, we solve the convex optimization problem (\ref{prob28}) to obtain the optimal time allocation $\{\tau_1, \tau_2\}$, and local computation frequency $f_k$. These steps are repeated until convergence is achieved. \textcolor{black}{The SCA and convex optimization steps  ensure that the objective value is non-decreasing at each iteration. Although PSO exhibits strong global search capability, it is inherently heuristic and cannot guarantee convergence to the global optimum within a finite number of iterations. To ensure monotonic improvement of the objective, we adopt an enumeration-based strategy that initializes the PSO algorithm with multiple starting points and selects the best result. Furthermore, the previous PA positions are retained if the updated ones do not lead to a higher objective value.}

\section{Simulation Results}
In this section, we present simulation results to validate the effectiveness of the proposed scheme and optimization method. Specifically, we compare the proposed scheme against the following three baseline schemes:
\begin{itemize}
\item \textbf{Conventional MIMO scheme}: A uniform linear array (ULA)  with half-wavelength spacing is deployed at location (0, 0, $d$) to serve $K$ devices. 
\item \textbf{Fixed PA deployment scheme}: The PAs are uniformly  distributed along the waveguide \textcolor{black}{and remain fixed throughout the operation.}
\item \textbf{TDMA scheme}: The entire frame duration is equally divided into $K+1$ time slots, where one slot is allocated to downlink WPT and the remaining $K$ slots are assigned to uplink task offloading for each device.
\end{itemize}

 For the simulation setup, we consider a rectangular area of size $D_x\times D_y=30~ \text{m}\times10~\text{m}$, with the waveguide height set to $d=3$ m. The total frame duration is $T=1$ s. We assume $K = 4$ devices and $M = 4$ PAs in the system. The BS transmit power is  $P_B=43$ dBm, the noise power spectral density is set to $-174$ dBm/Hz, and the signal bandwidth is $B=100$ MHz. The task computation intensity is set to $D=200$ cycles/bit. In addition, we set $f_c=28$ GHz, $n_e=1.4$ \cite{PASS2,PASS3}, and $\kappa=1e^{-28}$ \cite{WPT2}.

\begin{figure}[!t]
 	\centering
 	\includegraphics[width=3.3in]{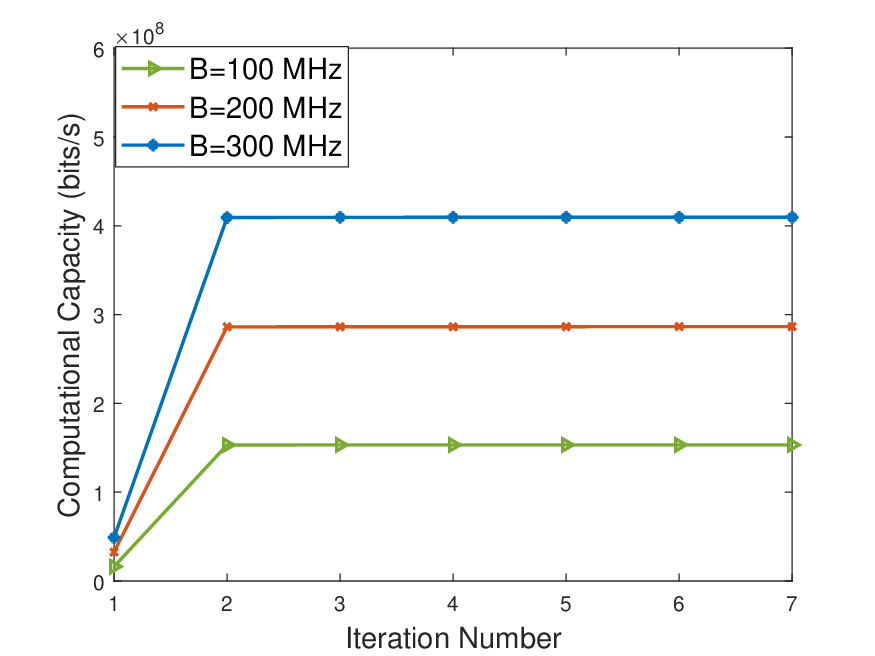}
 	\caption{The convergence performance of Algorithm 2.}
 	\label{fig:0}
 \end{figure}

We first evaluate the convergence behavior of the proposed Algorithm 2 under different signal bandwidths $B$, as illustrated in Fig.~{\ref{fig:0}}. It is observed that the alternating optimization algorithm exhibits rapid convergence across  all tested bandwidth levels. Furthermore, increasing $B$ \textcolor{black}{leads to an improved uplink offloading rate,  thereby resulting in  higher computation capacity.}
\begin{figure}[!t]
	\centering
	\includegraphics[width=3.3in]{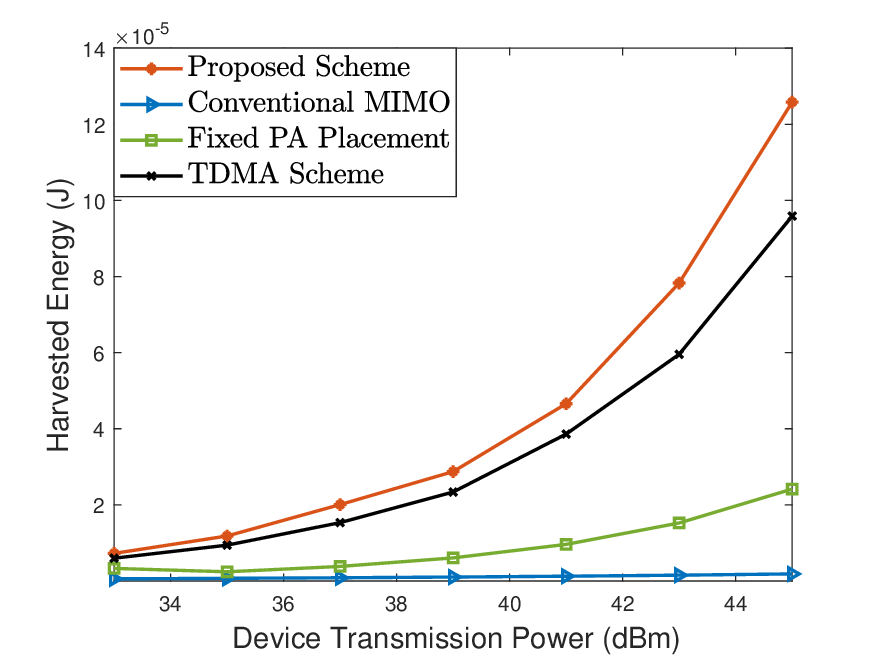}
	\caption{Harvested energy versus BS's transmit power $P_B$.}
	\label{fig:3}
\end{figure}

To illustrate the advantages of PAs, Fig.~{\ref{fig:3}} presents the harvested energy of different schemes during the downlink WPT phase. It can be observed that the proposed scheme \textcolor{black}{achieves the highest harvested energy among all evaluated designs.} This benefit stems from the joint optimization of PA positions and radiation factors, which effectively reduces the path loss from the BS to the devices. Moreover, we observe that the schemes with PA design exhibit steeper slopes,  \textcolor{black}{indicating that they are more sensitive to increases in BS transmit power and thus can better exploit higher transmission levels for energy harvesting.}

\begin{figure}[!t]
	\centering
	\includegraphics[width=3.3in]{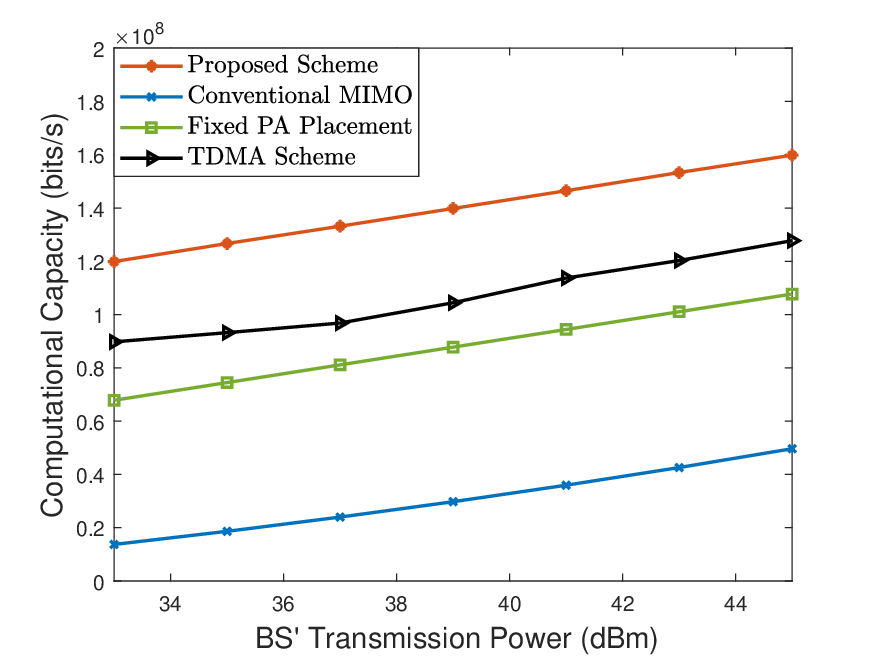}
	\caption{Computation capacity versus BS's transmit power $P_B$.}
	\label{fig:1}
\end{figure}

In Fig.~{\ref{fig:1}}, we evaluate the computation capacity of different schemes under varying BS transmit power $P_B$. With increasing base station transmit power $P_B$, all four schemes benefit from improved device energy harvesting, leading to enhanced overall performance. Compared to the conventional MIMO scheme, the other three schemes achieve notably higher computation capacities. \textcolor{black}{This improvement is primarily attributed to the ability of PAs to reduce the path loss between the BS and devices, in contrast to fixed-position ULAs,  thereby enhancing energy harvesting and task offloading efficiency.} Moreover, the proposed scheme  outperforms the fixed PA deployment scheme by flexibly optimizing PA positions to enhance system performance. Compared to the TDMA-based scheme, the proposed NOMA-based design more effectively exploits time-domain multiplexing, enabling increased task offloading and improved overall computation capacity.

\begin{figure}[!t]
	\centering
	\includegraphics[width=3.3in]{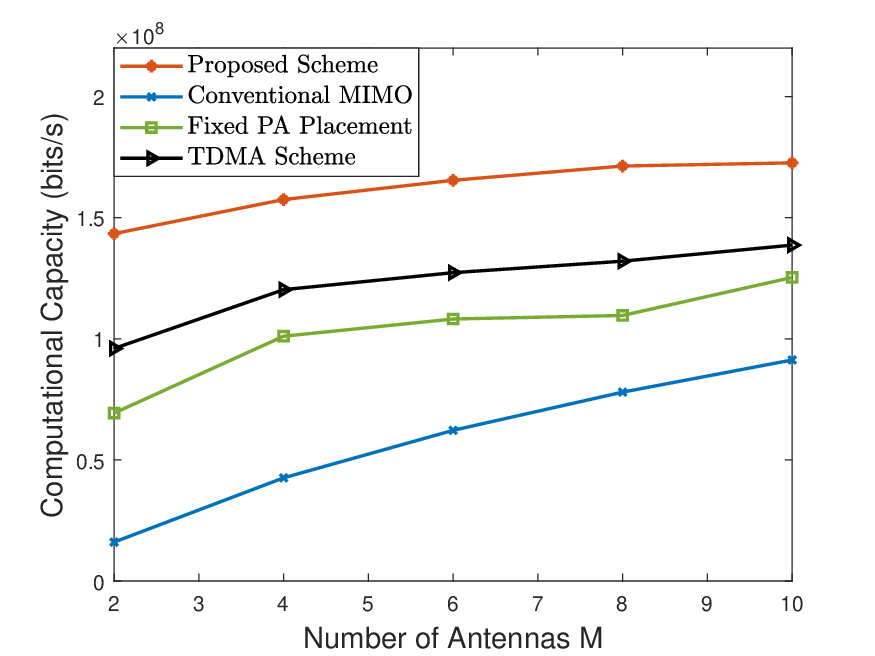}
	\caption{Computation capacity versus the  number of  antennas $M$.}
	\label{fig:2}
\end{figure}

In Fig.~{\ref{fig:2}}, we illustrate the variation in the  computation capacity with respect to the number of antennas $M$. The proposed scheme demonstrates superior performance across all antenna configurations, demonstrating its effectiveness. Additionally, as $M$ increases, all schemes show improved computation capacity, owing to the higher signal reception gain and greater spatial degrees of freedom enabled by additional antennas.

\section{Conclusion}
This paper proposed a novel integration of PAs into wireless powered MEC systems to simultaneously enhance energy harvesting and computational performance. By enabling spatially reconfigurable signal transmission and reception through the controllable dielectric elements along extended waveguides, the PA-assisted architecture effectively reduced large-scale path loss in both the downlink WPT and uplink task offloading phases. To tackle the resulting non-convex joint optimization involving time allocation, transmit power, and PA positioning, we proposed an alternating algorithm that combines PSO with SCA. Simulation results validated the effectiveness of the proposed design, demonstrating notable  performance gains over conventional antenna systems and highlighting the potential of PA-enabled MEC architectures for future low-power and computation-intensive IoT applications.


\bibliographystyle{IEEEtran}
\bibliography{biblp/bibfilelp}

\begin{thebibliography}{10}
\providecommand{\url}[1]{#1}
\csname url@samestyle\endcsname
\providecommand{\newblock}{\relax}
\providecommand{\bibinfo}[2]{#2}
\providecommand{\BIBentrySTDinterwordspacing}{\spaceskip=0pt\relax}
\providecommand{\BIBentryALTinterwordstretchfactor}{4}
\providecommand{\BIBentryALTinterwordspacing}{\spaceskip=\fontdimen2\font plus
\BIBentryALTinterwordstretchfactor\fontdimen3\font minus
  \fontdimen4\font\relax}
\providecommand{\BIBforeignlanguage}[2]{{%
\expandafter\ifx\csname l@#1\endcsname\relax
\typeout{** WARNING: IEEEtran.bst: No hyphenation pattern has been}%
\typeout{** loaded for the language `#1'. Using the pattern for}%
\typeout{** the default language instead.}%
\else
\language=\csname l@#1\endcsname
\fi
#2}}
\providecommand{\BIBdecl}{\relax}
\BIBdecl

\bibitem{iot}
D.~C. Nguyen, M.~Ding, P.~N. Pathirana, A.~Seneviratne, J.~Li, D.~Niyato,
  O.~Dobre, and H.~V. Poor, ``{6G} internet of things: A comprehensive
  survey,'' \emph{IEEE Internet Things J.}, vol.~9, no.~1, pp. 359--383, Jan.
  2022.

\bibitem{MEC-WPT}
F.~Wang, J.~Xu, and S.~Cui, ``Optimal energy allocation and task offloading
  policy for wireless powered mobile edge computing systems,'' \emph{IEEE
  Trans. Wireless Commun.}, vol.~19, no.~4, pp. 2443--2459, Apr. 2020.

\bibitem{RIS1}
Q.~Wu and R.~Zhang, ``Intelligent reflecting surface enhanced wireless network
  via joint active and passive beamforming,'' \emph{IEEE Trans. Wireless
  Commun.}, vol.~18, no.~11, pp. 5394--5409, Nov. 2019.

\bibitem{RIS2}
M.~Hua, Q.~Wu, W.~Chen, Z.~Fei, H.~C. So, and C.~Yuen, ``Intelligent reflecting
  surface-assisted localization: Performance analysis and algorithm design,''
  \emph{IEEE Wireless Commun. Lett.}, vol.~13, no.~1, pp. 84--88, Jan. 2024.

\bibitem{RIS3}
G.~Chen, Q.~Wu, W.~Chen, D.~W.~K. Ng, and L.~Hanzo, ``{IRS}-aided wireless
  powered {MEC} systems: {TDMA} or {NOMA} for computation offloading?''
  \emph{IEEE Trans. Wireless Commun.}, vol.~22, no.~2, pp. 1201--1218, Feb.
  2023.

\bibitem{fas1}
K.-K. Wong, A.~Shojaeifard, K.-F. Tong, and Y.~Zhang, ``Fluid antenna
  systems,'' \emph{IEEE Trans. Wireless Commun.}, vol.~20, no.~3, pp.
  1950--1962, Mar. 2021.

\bibitem{fas2}
K.-K. Wong and K.-F. Tong, ``Fluid antenna multiple access,'' \emph{IEEE Trans.
  Wireless Commun.}, vol.~21, no.~7, pp. 4801--4815, July 2022.

\bibitem{Mov1}
L.~Zhu, W.~Ma, and R.~Zhang, ``Movable antennas for wireless communication:
  Opportunities and challenges,'' \emph{IEEE Commun. Mag.}, vol.~62, no.~6, pp.
  114--120, June 2024.

\bibitem{Mov2}
L.~Zhu \emph{et~al.}, ``Modeling and performance analysis for movable antenna
  enabled wireless communications,'' \emph{IEEE Trans. Wireless Commun.},
  vol.~23, no.~6, pp. 6234--6250, June 2024.

\bibitem{Mov3}
Y.~Gao, Q.~Wu, and W.~Chen, ``Joint transmitter and receiver design for movable
  antenna enhanced multicast communications,'' \emph{IEEE Transactions on
  Wireless Communications}, vol.~23, no.~12, pp. 18\,186--18\,200, Dec. 2024.

\bibitem{suzuki2022pinching}
H.~O.~Y. Suzuki and K.~Kawai, ``Pinching antenna: Using a dielectric waveguide
  as an antenna,'' \emph{NTT DOCOMO Technical J}, vol.~23, no.~3, pp. 5--12,
  Jan. 2022.

\bibitem{PASS0}
Z.~Ding, R.~Schober, and H.~Vincent~Poor, ``Flexible-antenna systems: A
  pinching-antenna perspective,'' \emph{IEEE Trans. Commun.}, 2025, early
  Access, 2024, 10.1109/TCOMM.2025.3555866.

\bibitem{PASS4}
C.~Ouyang, Z.~Wang, Y.~Liu, and Z.~Ding, ``Array gain for pinching-antenna
  systems ({PASS}),'' \emph{IEEE Commun. Lett.}, 2025, early Access, 2025,
  10.1109/LCOMM.2025.3566299.

\bibitem{PASS1}
\BIBentryALTinterwordspacing
Y.~Liu, Z.~Wang, X.~Mu, C.~Ouyang, X.~Xu, and Z.~Ding, ``Pinching-antenna
  systems ({PASS}): Architecture designs, opportunities, and outlook,'' 2025.
  [Online]. Available: \url{https://arxiv.org/abs/2501.18409}
\BIBentrySTDinterwordspacing

\bibitem{PASS2}
Y.~Xu, Z.~Ding, and G.~K. Karagiannidis, ``Rate maximization for downlink
  pinching-antenna systems,'' \emph{IEEE Wireless Commun. Lett.}, vol.~14,
  no.~5, pp. 1431--1435, May 2025.

\bibitem{PASS3}
S.~A. Tegos, P.~D. Diamantoulakis, Z.~Ding, and G.~K. Karagiannidis, ``Minimum
  data rate maximization for uplink pinching-antenna systems,'' \emph{IEEE
  Wireless Commun. Lett.}, vol.~14, no.~5, pp. 1516--1520, May 2025.

\bibitem{WPT2}
B.~Li, J.~Liao, W.~Wu, and Y.~Li, ``Intelligent reflecting surface assisted
  secure computation of wireless powered {MEC} system,'' \emph{IEEE Trans. Mob.
  Comput.}, vol.~23, no.~4, pp. 3048--3059, Apr. 2024.

\bibitem{PSO}
M.~Clerc and J.~Kennedy, ``The particle swarm - explosion, stability, and
  convergence in a multidimensional complex space,'' \emph{IEEE Trans. Evol.
  Comput.}, vol.~6, no.~1, pp. 58--73, Feb. 2002.

\bibitem{2004Convex}
S.~Boyd and L.~Vandenberghe, \emph{Convex Optimization}.\hskip 1em plus 0.5em
  minus 0.4em\relax Cambridge University Press, 2004.

\end{thebibliography}

\end{document}